%% file: main.tex
\documentclass[10pt,twocolumn,letterpaper]{article}

\usepackage[pagenumbers]{cvpr} 

\input{preamble}

\definecolor{cvprblue}{rgb}{0.21,0.49,0.74}
\usepackage[pagebackref,breaklinks,colorlinks,citecolor=cvprblue]{hyperref}

\def\paperID{****} 
\def\confName{CVPR}
\def\confYear{2024}

\title{3D Paintbrush: Local Stylization of 3D Shapes with Cascaded Score Distillation}

\author{Dale Decatur\\
University of Chicago\\
\and
Itai Lang\\
University of Chicago\\
\and
Kfir Aberman\\
Snap Research\\
\and
Rana Hanocka\\
University of Chicago\\
}

\begin{document}
\twocolumn[{%
\renewcommand\twocolumn[1][]{#1}%
\maketitle
\input{figures/overhead/teaser}
}]
\input{sec/0_abstract}    
\input{sec/1_intro}
\input{sec/2_related_work}
\input{sec/3_method}

\input{sec/4_experiments}
\input{sec/5_conclusion}
{
    \small
    \bibliographystyle{ieeenat_fullname}
    \bibliography{main}
}


\end{document}

%% file: preamble.tex
\usepackage[dvipsnames]{xcolor}
\usepackage{overpic}
\usepackage{amsmath}
\usepackage{algorithm}
\usepackage{algpseudocode}
\usepackage{float}

\newif\ifdraft
\draftfalse

\ifdraft
\newcommand{\rana}[1]{{\color{blue}[\textbf{Rana:} #1]}}
\newcommand{\dale}[1]{{\color{red}[\textbf{Dale:} #1]}}
\newcommand{\itai}[1]{{\color{green}[\textbf{Itai:} #1]}}
\newcommand{\kfir}[1]{{\color{magenta}[\textbf{Kfir:} #1]}}

\else
\newcommand{\rana}[1]{}
\newcommand{\dale}[1]{}
\newcommand{\itai}[1]{}
\newcommand{\kfir}[1]{}

\fi

\newcommand{\ourmethod}{3D Paintbrush}


%% file: figures/overhead/teaser.tex
\begin{center}
    \centering
    \newcommand{\pl}{-2}
    \begin{overpic}[width=\textwidth, trim=0 0 0 40]{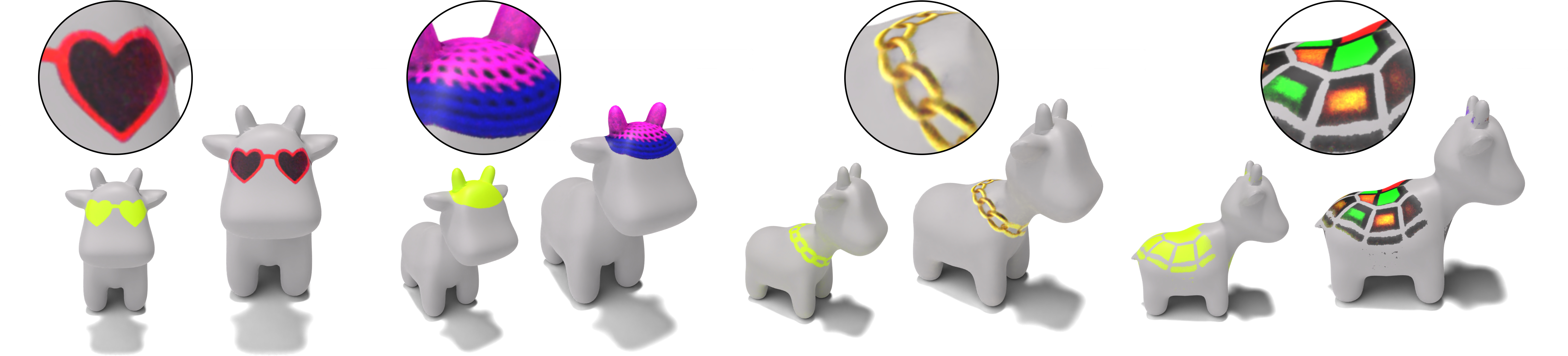}
    \put(3,  \pl){\textcolor{black}{Heart shaped sunglasses}}
    \put(27,  \pl){\textcolor{black}{Colorful crochet hat}}
    \put(51,  \pl){\textcolor{black}{Gold chain necklace}}
    \put(75,  \pl){\textcolor{black}{Stained glass turtle shell}}
    \end{overpic}
    \vspace{-2mm}
    \captionof{figure}{Utilizing only a text prompt as guidance, 3D Paintbrush seamlessly generates local stylized textures on bare meshes. Our approach produces a localization map (yellow regions) and a highly detailed texture map which conforms to it.}
    \label{fig:teaser}
\end{center}

%% file: sec/0_abstract.tex
\begin{abstract}
In this work we develop \ourmethod{}, a technique for automatically texturing local semantic regions on meshes via text descriptions. Our method is designed to operate directly on meshes, producing texture maps which seamlessly integrate into standard graphics pipelines. We opt to simultaneously produce a localization map (to specify the edit region) and a texture map which conforms to it. This synergistic approach improves the quality of both the localization and the stylization. To enhance the details and resolution of the textured area, we leverage multiple stages of a cascaded diffusion model to supervise our local editing technique with generative priors learned from images at different resolutions. Our technique, referred to as Cascaded Score Distillation (CSD), simultaneously distills scores at multiple resolutions in a cascaded fashion, enabling control over both the granularity and global understanding of the supervision. We demonstrate the effectiveness of \ourmethod{} to locally texture a variety of shapes within different semantic regions. Project page: \url{https://threedle.github.io/3d-paintbrush}

\end{abstract}

%% file: sec/1_intro.tex
\section{Introduction}
\label{sec:intro}

\input{figures/overhead/composite}

The ability to edit existing high-quality 3D assets is a fundamental capability in 3D modeling workflows. Recent works have shown exceptional results for text-driven 3D data creation \cite{ruiz2023dreambooth, sjc, lin2023magic3d, tsalicoglou2023textmesh, metzer2022latent, wang2023prolificdreamer}, but focus on making \textit{global} edits. While some progress has been made on local editing using an explicit localization of the edit region~\cite{sella2023voxe, zhuang2023dreameditor}, these regions are often coarse and lack fine-grained detail. Highly-detailed and accurate localizations are important for constraining the edits to be within a specific region, preventing changes unrelated to the target edit. Furthermore, while meshes with texture maps are the de facto standard in graphics pipelines, existing local editing work does not natively operate on meshes nor produce texture maps for them.

In this work we develop \ourmethod{}, a method for automatically texturing local semantic regions on meshes via text descriptions. Our method is designed to operate directly on meshes, producing texture maps which seamlessly integrate into standard graphics pipelines. 
\ourmethod{} is controlled via intuitive, free-form text input, allowing users to describe their edits using open vocabulary on a wide range of meshes. Specifically, given an input mesh and a text prompt, \ourmethod{} produces the corresponding high-quality texture map and a localization region to confine it. To enhance the details and resolution of the local textured area, we introduce Cascaded Score Distillation (CSD) which leverages multiple stages of a cascaded diffusion model. Our explicit localization masks can be used to layer our edit texture onto existing textures. More generally, we may perform other edits beyond texturing such as modifying the localization region (\eg translating the texture along the surface), or deforming the geometry within the localization region.

We opt to represent both our localization map and texture map as neural fields encoded by multi-layer perceptions (MLPs). This formulation enables \emph{triangulation-agnostic} and super-resolution localization and texture maps. 
Our method synthesizes both a fine-grained localization mask and high-quality texture in tandem. Simultaneously generating the localization and texture maps improves the quality of each. 
The texture map drives the localization to become more detailed and intricate. 
The localization explicitly masks the texture, ensuring a coherent local style which respects the localization boundary. 

Our local stylization operates in small regions, necessitating higher resolution supervision compared to global generative techniques. Existing approaches leverage pretrained text-to-image diffusion models with Score Distillation Sampling (SDS) to supervise text-driven optimizations \cite{li2023focaldreamer, sjc}. Text-to-image diffusion models often contain multiple cascaded stages in order to achieve high resolution \cite{ho2022cascaded}, but standard SDS only utilizes the first low-resolution stage of the cascaded model. Our technique, referred to as Cascaded Score Distillation (CSD), simultaneously distills scores at multiple resolutions in a cascaded fashion, enabling control over both the granularity and global understanding of the supervision. Since cascaded stages are trained entirely independently, our insight is to formulate a distillation loss that incorporates all stages in tandem.

In summary, our method enables local text-driven stylization of meshes. By explicitly learning a localization in tandem with the texture, we ensure that our edits are bounded by the localized region. Due to the synergistic optimization, \ourmethod{}'s localizations are sharper than those produced by existing methods such as \cite{abdelreheem2023SATR, decatur2023highlighter, zhuang2023dreameditor, sella2023voxe}. Using our CSD, which leverages all stages of the diffusion model, we can control the granularity and global understanding of the supervision achieving higher resolution textures and localizations than standard SDS. We demonstrate that \ourmethod{} yields diverse local texturing on a variety of shapes and semantic regions.

%% file: figures/overhead/composite.tex
\begin{figure*}
    \centering
    \newcommand{\pl}{-2}
    \begin{overpic}[width=\linewidth]{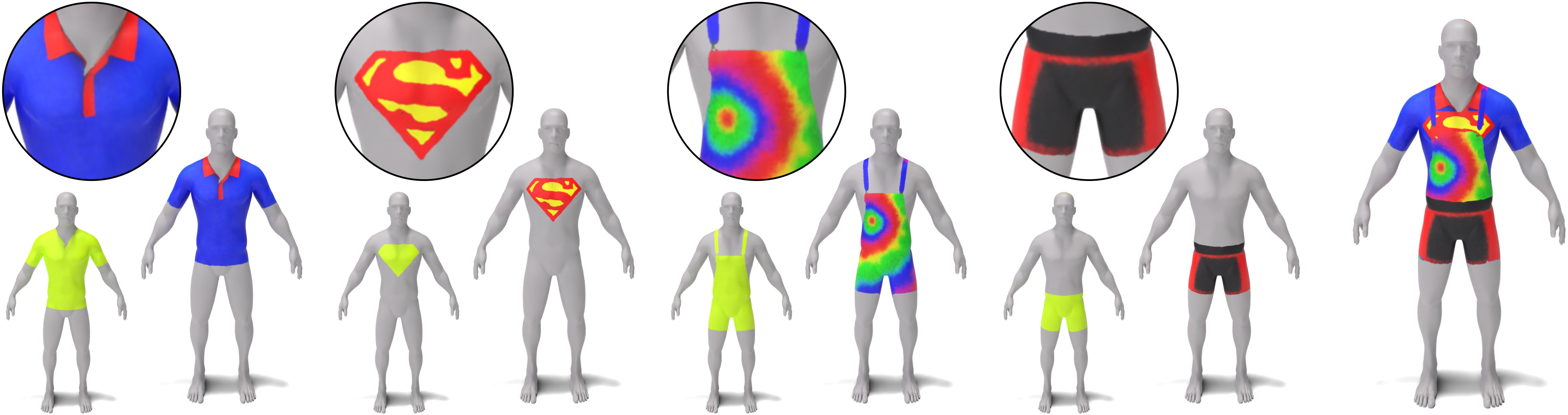}
    \put(2,  \pl){\textcolor{black}{Colorful polo shirt}}
    \put(23,  \pl){\textcolor{black}{Superman emblem}}
    \put(46,  \pl){\textcolor{black}{Tie-dye apron}}
    \put(66,  \pl){\textcolor{black}{Muay Thai shorts}}
    \put(89,  \pl){\textcolor{black}{Composite}}
    \end{overpic}
    \vspace{-2mm}
    \caption{\textbf{Precise composition of multiple local textures}. \ourmethod{} produces highly-detailed textures that effectively adhere to the predicted localizations. This enables seamlessly compositing local textures without unwanted fringes (right).}
    \label{fig:composite}
\end{figure*}
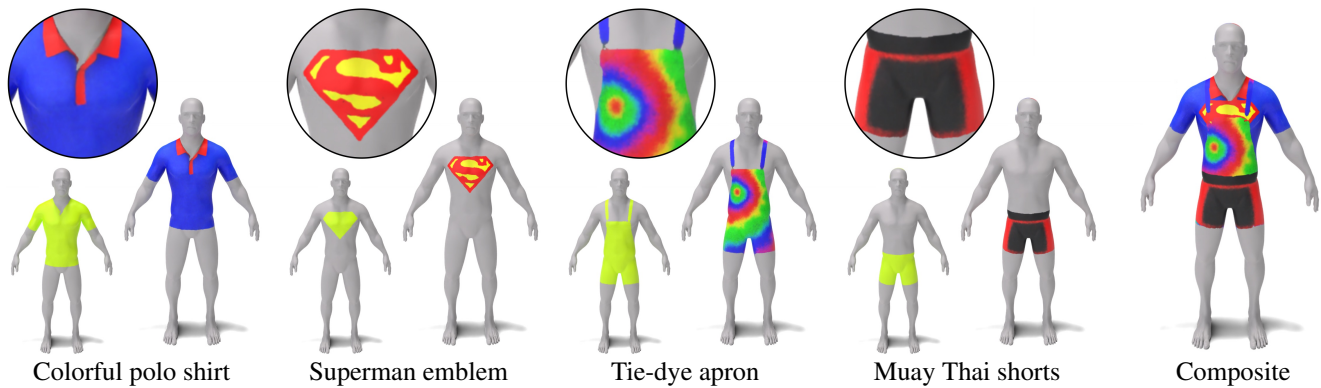

%% file: sec/2_related_work.tex
\section{Related Work}
\label{sec:related_work}
\noindent \textbf{Neural styles.} Existing work uses neural networks~\cite{michel2022text2mesh, oechsle2019texture, bokhovkin2023mesh2tex, siddiqui2022texturify, huang2020adversarial, wei2021deep, mohammad2022clip, metzer2022latent, ma2023x, lei2022tango, hertz2020deep, Liu:Subdivision:2020, yin20213dstylenet, hollein2022stylemesh, gao2021tm} for mesh stylization. Some methods opt for a neural field over the mesh vertices~\cite{michel2022text2mesh} which allows for triangle agnostic styles, while others optimize a texture map~\cite{mohammad2022clip, yin20213dstylenet}. Other works use a neural radiance field NeRF~\cite{mildenhall2020nerf} for stylization~\cite{zhang2022arf, liu2023stylerf, fan2022unified}. Yet, these works focus on stylization rather than localization.

\input{figures/overhead/3dp_overview}
\smallskip
\noindent \textbf{Text-driven analysis.} Large 2D models have been used for analytical tasks in 3D such as localization and segmentation~\cite{decatur2023highlighter, abdelreheem2023SATR, kerr2023lerf, kobayashi2022distilledfeaturefields, tschernezki2022neural, vora2021nesf, ha2022semabs, zhuang2023dreameditor}. 3D Highlighter~\cite{decatur2023highlighter} localizes a text-driven region on a 3D shape using CLIP's~\cite{radford2021learning} joint image-language embedding space as guidance to supervise the optimization. SATR~\cite{abdelreheem2023SATR} uses GLIP~\cite{li2022grounded}, a text-driven 2D segmentation model, and directly projects the segmentations back to the 3D shape. All of these works aim to generate localizations, but do not produce textures. Furthermore, only \cite{decatur2023highlighter, abdelreheem2023SATR, zhuang2023dreameditor} aim to produce a tight localization on meshes and we find that these approaches still produce relatively smooth localization regions that cannot capture the high frequency details needed for sharp local edits.

\smallskip
\noindent \textbf{Text-driven generation and editing.}
Existing works have leveraged pre-trained 2D models to generate 3D representations that adhere to a text prompt~\cite{lee2022understanding, gao2023textdeformer, jain2022zero, michel2022text2mesh, fu2022shapecrafter, mohammad2022clip, zeng2022lion, babu2023hyperfields}. Recent methods do so with text-conditioned 2D diffusion models by distilling scores~\cite{poole2022dreamfusion, sjc} from the 2D model onto the 3D representation~\cite{poole2022dreamfusion, sjc, zhu2023hifa, tsalicoglou2023textmesh, lin2023magic3d, chen2023fantasia3d, shi2023mvdream, tsalicoglou2023textmesh}. Instead of generating both geometry and styles from scratch, some works optimize the texture of an existing, fixed geometry~\cite{metzer2022latent, oechsle2019texture, chen2023text2tex, michel2022text2mesh}. Further, another line of works aims to generate 3D representations from images~\cite{gao2022get3d, liu2023one, qian2023magic123, liu2023zero1to3}. Different from our objective, these works aim to generate or globally manipulate existing 3D representations, while our work focuses on local editing. Existing text-to-3D methods typically used for generation~\cite{poole2022dreamfusion, metzer2022latent, sjc, wang2023prolificdreamer} can also be used to perform global edits by cleverly designing text prompts and using negative prompting. Applying DreamBooth~\cite{ruiz2023dreambooth} fine tuning to these methods can further improve editing performance~\cite{zhuang2023dreameditor}. Instruct-NeRF2NeRF~\cite{haque2023instruct} explicitly performs 3D editing using pretrained image editing methods. However, all of these approaches have no explicit localization for their edits. Without explicit edit regions, there are no guarantees that edits will be contained to the desired region specified (or implied) by the text. Thus, these methods struggle to perform highly specific local edits without changing other components of the 3D representation's appearance.

\smallskip
\noindent \textbf{Text-driven local editing.} While many approaches can perform global edits, few works have addressed the task of precise, local editing for 3D representations. Local editing is challenging since, in addition to synthesizing the edit, methods need to localize the edit region. Progress has been made on local editing in images and videos~\cite{bar2022text2live, brooks2023instructpix2pix, gal2022image, hertz2022prompt}. In 3D, FocalDreamer~\cite{li2023focaldreamer} addresses region localization by requiring user-specified elliptical edit regions. This gives precise edit regions, but requires additional, tedious user input compared to strictly text-driven approaches. Vox-E~\cite{sella2023voxe} works with voxel representations, first generating a global edit and then using a cross-attention map between the original and edited objects to mask the edited object. This gives an explicit localization, but does not ensure that the edit conforms the bounds of the localization since the localization mask is determined after the edit is generated. DreamEditor~\cite{zhuang2023dreameditor} explicitly localizes the edit region using a text-to-image model's attention maps to restrict the edits to NeRF regions that are salient when conditioned on the text-specified edit. Since both Vox-E and DreamEditor use attention maps to define their edit regions, there is no incentive for the localization to be visually meaningful in isolation and thus these methods produce coarse localizations that are often larger than the actual edit. This can lead to undesired changes to the appearance of the object in regions outside of intended edit region. Our approach imposes a visual loss on our localizations in order to enforce sharp boundaries that are tightly coupled with our texture edits. Additionally, since existing purely text-driven local editing approaches only work on voxels and NeRFs, our approach is the first to enable text-driven local editing on meshes.

\smallskip
\noindent \textbf{High resolution text-to-3D}.
Several works have explored techniques to increase the resolution for text-to-3D. Many recent works apply SDS to latent diffusion models~\cite{sjc, metzer2022latent, zhu2023hifa, lin2023magic3d, wang2023prolificdreamer}. Typical latent score distillation approaches apply the SDS gradient in latent space~\cite{sjc, metzer2022latent}. However recent works backpropagate the gradient through the encoder to get gradients in higher resolution 512x512 RGB space~\cite{lin2023magic3d, wang2023prolificdreamer, zhu2023hifa}. Dreamtime~\cite{huang2023dreamtime} and ProflicDreamer~\cite{wang2023prolificdreamer} have also explored altering the timestep annealing to give less noisy supervision towards the end of the optimization and increase the detail of their generations. HiFA~\cite{zhu2023hifa} proposes BGT+ loss which each iteration denoises over multiple successive timesteps to provide better gradients and achieve high fidelity appearance. While all of these approaches have shown impressive improvements to the resolution of SDS supervision, SDS only utilizes the base stage (not super-resolution stages). Thus, these proposed improvements are orthogonal to ours -- they can be incorporated at the super-resolution stages using CSD as well.

%% file: figures/overhead/3dp_overview.tex
\begin{figure*}[t]
    \centering
    \includegraphics[width=\linewidth]{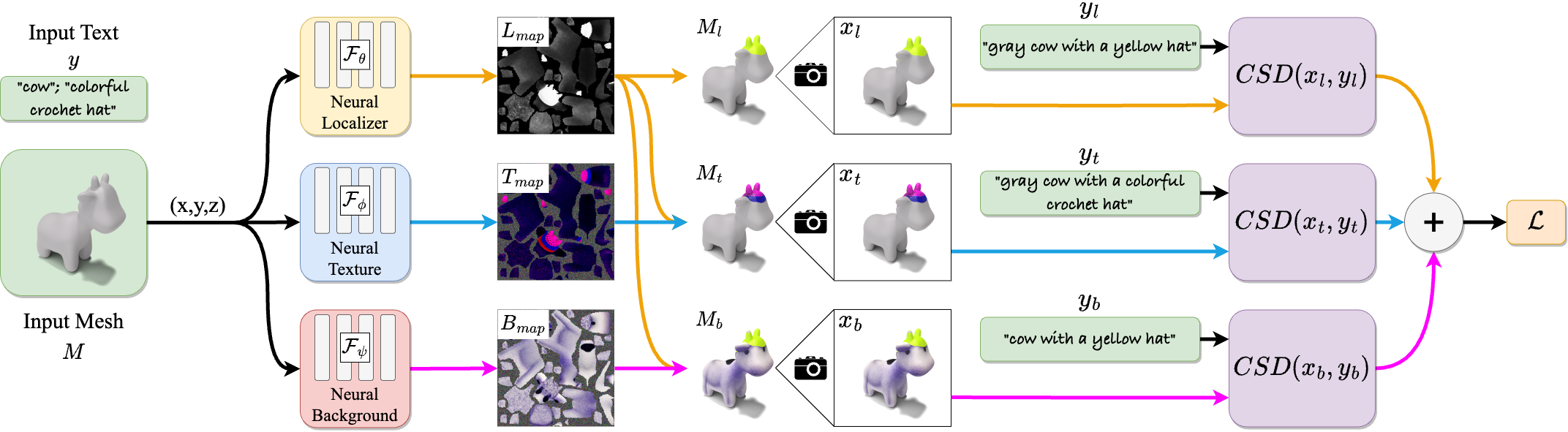}
    \caption{\textbf{Overview of \ourmethod{}}. Each point on the surface of the mesh is passed into three different branches to produce a localization probability, texture map, and background map. We texture three different variants of the same mesh with the localization, texture, and background maps and render them from the same viewpoint. Each image along with the corresponding text condition is used to compute the CSD loss.}
    \label{fig:overview}
\end{figure*}

%% file: sec/3_method.tex
\input{figures/overhead/gallery}

\section{Method}
\label{sec:method}

We show an overview of our method in \cref{fig:overview}. The inputs to our system are a mesh $M$ and a text description $y$ of the desired local edit. Our system produces a local texture on the mesh $M$ that adheres to the text prompt $y$. To supervise our optimization, we use score distillation with a pretrained text-to-image diffusion model. However, local editing requires higher detail than standard generation due to the small size and granularity of the desired edits. In order to further improve the detail of our localization and texture, we introduce Cascaded Score Distillation (CSD), a technique that distills scores at multiple resolutions of the 2D cascaded model. This approach enables leveraging all stages of a cascaded model and provides control over both the detail and global understanding of the supervision.

\subsection{Local Neural Texturing}
\ourmethod{} represents local textures as neural texture maps over the surface of a mesh $M$ defined by vertices $V \in \mathbb{R}^{n \times 3}$ and faces $F \in \{1,...,n\}^{m \times 3}$. Extracting an explicit texture map from our neural textures is trivial, making our representation compatible with existing graphics pipelines. Furthermore, using texture maps enables producing high resolution textures (i.e. sub-triangle values) without a computationally expensive high resolution mesh. 
A straight-forward approach of directly optimizing texture values results in texture maps with artifacts and noise (see supplemental material). To mitigate this, we leverage the smoothness of neural networks~\cite{spectralbias}. However, a straight-forward application of an MLP to a 2D texture map ($(u, v) \rightarrow (r, g, b)$) is inherently invalid at the texture seams (\eg erroneous interpolations at boundaries), which may lead to texture discontinuities on the rendered mesh.

Thus, we instead formulate our MLPs to operate on 3D coordinates leading to predictions in 3D that are inherently smooth and without any seam discontinuities. To do so, we invert the UV mapping $\psi(x,y,z)=(u,v)$ to get a map $\psi^{-1}(u,v)=(x,y,z)$ from 2D texels to 3D coordinates on the surface of the mesh. We optimize our MLPs with the 3D coordinates obtained from the 2D texel centers. We employ two primary networks, one for localization and one for texturing. Our neural \textit{localization} MLP is a function $\mathcal{F_{\theta}}$ that maps a 3D coordinate $\mathbf {x} = (x,y,z)$ to a probability $p$ (which we map back to a 2D localization map). Similarly, our neural \textit{texture} MLP is a function $\mathcal{F_{\phi}}$ that takes in a 3D coordinate and outputs an RGB value $R, G, B$ (which we map back to a 2D texture image). Our architecture first passes the 3D coordinates through positional encoding~\cite{tancik2020fourfeat} before going through a 6-layer MLP. This formulation of using MLPs defined on the 3D surface leads to a neural texture which produces smoothly varying outputs in 3D, even though our 2D texture maps have discontinuities at the texture seams. The smoothness provided by the MLPs reduces artifacts, produces less noisy textures, and provides super resolution capabilities. Although we optimize our MLPs with 3D coordinates mapped from 2D texel centers, during inference, we may query the MLP for any value (\ie sub-texels that enable super resolution texture maps even across seams).

\input{figures/overhead/simultaneous}

\subsection{Visual Guidance for Localized Textures}
\label{subsec:visual-guidance}
We guide our optimization using three distinct losses that encourage both the localization and texture map towards visually desirable results. Each loss is visualized as a branch in \cref{fig:overview} -- top branch: localization loss, middle branch: local texture map loss, bottom branch: background loss.

\smallskip
\noindent
\textbf{Local texture map loss.} First, we obtain our localization map $L_{map} \in \mathbb{R}^{H \times W}: 0 \leq L_{map} \leq 1$ from the neural localization MLP $L_{map} = \psi(\mathcal{F_{\theta}}(\mathbf{x}))$ and the texture map $T_{map} \in \mathbb{R}^{H \times W}$ from the neural texture MLP $T_{map} = \psi(\mathcal{F_{\phi}}(\mathbf{x}))$.
We use the localization $L_{map}$ to mask the texture $T_{map}$ to get a local texture map $T_{map}'$ which only contains textures inside the localization region.
We apply the masked texture $T_{map}'$ to our mesh $M$ to get a locally-textured mesh $M_t$ and construct a local-texture text prompt $y_t$ from the input text $y$ (middle branch \cref{fig:overview}). We then supervise our optimization using a text-conditioned visual loss (cascaded score distillation, see \cref{subsec:csd}) on $M_t$ and $y_t$. By applying a visual loss to the localization-masked texture, we get informative and meaningful gradients for both our texture MLP and our localization MLP.

\smallskip
\noindent
\textbf{Localization loss.}
Using only the texture loss allows for trivial solutions where the mask contains a region that includes, but is much larger than, the desired localization region. To encourage the localization region to be meaningful, we employ a visual loss on the localization region in isolation (similar to 3D Highlighter~\cite{decatur2023highlighter}). Specifically, we blend a (yellow) color onto the mesh according to the localization map to get a localization-colored mesh $M_l$ (top branch \cref{fig:overview}). From the text input $y$, we derive a target localization prompt $y_l$ describing the localized region in the format used in 3D Highlighter~\cite{decatur2023highlighter}. We then use $M_l$ and $y_l$ as input to the text-conditioned visual loss. Using this loss significantly improves the detail and quality of the localization.

\smallskip
\noindent
\textbf{Background loss.} Using only the top two branches in \cref{fig:overview} leads to broader localizations that incorporate superfluous elements characteristic of the input 3D model (\ie a bill on a duck), in addition to the desired localization region (see \cref{fig:background-ablation}). To mitigate this, we learn a background texture $B_{map} \in \mathbb{R}^{H \times W}$ that intentionally contains these characteristic elements of the input 3D shape in the inverse of the localization region $1 - L_{map}$ (the area outside the localization region). Specifically, we blend both the background texture $B_{map}$ (using $1 - L_{map}$) and a yellow color (using $L_{map}$) to get a composited texture $B_{map}' = L_{map}(\textsc{yellow}) + (1-L_{map})B_{map}$ (bottom branch in \cref{fig:overview}). We apply the composited texture $B_{map}'$ to the mesh to get $M_b$ and then supervise the background MLP using a visual loss conditioned on both $M_b$ and a target text $y_b$ (derived from $y$). The target text $y_b$ describes the generic object class (\ie `cow' in \cref{fig:overview}) with a colored (yellow) localization region. See supplemental material for more details. The third loss directly encourages incorporating the superfluous elements in the background texture which \textit{discourages} the localization region from incorporating such undesired elements (since $L_{map}$ and $1-L_{map}$ are mutually exclusive). 

Key to our method is the simultaneous optimization of the localization map (that specifies the edit region) \textit{and} the texture map that conforms to it. This approach improves the quality of both the localization and the stylization. The texture map drives the localization to become more detailed and intricate, while the localization explicitly masks the texture, ensuring a coherent local style which respects the localization boundary (see \cref{fig:simultaneous}).

\subsection{Score Distillation and Cascaded Diffusion}
\label{subsec:sds}
\noindent \textbf{Score Distillation.}
To guide our local stylization, we leverage powerful pretrained text-to-image diffusion models. Existing approaches use these models in conjunction with Score Distillation Sampling (SDS) to supervise text-driven optimizations \cite{poole2022dreamfusion, sjc}. For each iteration of an optimization of an image $x$ that we want to supervise with diffusion model $\phi$ and text prompt $y$, SDS \cite{poole2022dreamfusion} proposes the following gradient:
\begin{equation}
    \nabla_x \mathcal{L}_{SDS}(\phi, x, y) = w(t)(\epsilon_{\phi}(z_t, t, y) - \epsilon)
    \label{eq:sds}
\end{equation}
where timestep $t \sim \mathcal{U}(\{1,\ldots,T\})$ is sampled uniformly and noise $\epsilon \sim \mathcal{N}(\mathbf{0},\mathbf{I})$ is Gaussian. The noisy image $z_t$ is obtained by applying a timestep-dependent scaling of $\epsilon$ to the image $x$. The weight $w(t)$ is a timestep-dependent weighting function and $\epsilon_{\phi}(z_t, t, y)$ is the noise predicted by the diffusion model conditioned on $z_t$, $t$, and $y$. Note that \cref{eq:sds} omits the U-Net Jacobian term that \cite{poole2022dreamfusion} find is not needed in practice. This objective is similar to the objective used in diffusion model training, however, instead of optimizing the weights of the model, this gradient is applied to the image $x$.

\smallskip
\noindent \textbf{Cascaded Diffusion.} 
Text-to-image diffusion models often contain multiple cascaded stages at different resolutions in order to achieve high resolution outputs~\cite{ho2022cascaded}. These cascaded diffusion models consist of a base stage $\phi^1$ (stage $1$) and some number of super-resolution stages $\phi^{i>1}$ (stages $2$-$N$). The base stage is identical to a standard diffusion model, predicting noise $\epsilon_{\phi^1}(z^1_t, t, y)$ conditioned on noisy image $z^1_t$, timestep $t$, and text prompt $y$. However, the super-resolution stages are conditioned on two differently-noised images: one at the current resolution ($z^{i}_t$ with timestep $t$ and noise $\epsilon^i$) and one at the lower resolution ($z^{i-1}_s$ with timestep $s$ and noise $\epsilon^{i-1}$). The predicted noise for the super resolution stage is given by $\epsilon_{\phi^i}(z^i_t, t, z^{i-1}_s, s, y)$. During inference, the lower resolution input image is obtained by adding noise to the output of the prior stage. However during training, both the high and low resolution images are obtained by sampling a single image from the training dataset and rescaling it to different resolutions.

Standard SDS~\cite{poole2022dreamfusion} only utilizes the first low-resolution base stage, thus neglecting the full potential of the cascaded model. It is not immediately obvious how to formulate a score distillation technique for all stages of a cascaded diffusion model since super-resolution stages take multiple resolution inputs and, at inference, they require a fully denoised output from the prior stage~\cite{ho2022cascaded}. We take inspiration from SDS and use the perspective of \textit{diffusion training} as opposed to inference, and extend it to the training of cascaded diffusion models. As far as we can ascertain, we are the first to consider score distillation using the cascaded super-resolution stages.

\input{figures/overhead/csd_overview}
\subsection{Cascaded Score Distillation}
\label{subsec:csd}

\noindent \textbf{CSD overview.} Our technique, referred to as Cascaded Score Distillation (CSD), simultaneously distills scores at multiple resolutions in a cascaded fashion (illustrated in \cref{fig:csd-overview}). Since the stages of a cascaded diffusion model $\phi$ are trained entirely independently of one another, our insight is to formulate a distillation loss that incorporates gradients from all stages $(\phi^1,...,\phi^N)$ simultaneously. We observe that different stages of the cascaded model provide different levels of granularity and global understanding (\cref{fig:csd-ablation}). Controlling the influence of each stage provides control over the details and the corresponding localization of the supervision (\cref{fig:csd-control}).

\input{figures/overhead/csd_ablation}

\input{figures/overhead/csd_control}

\smallskip
\noindent \textbf{CSD Formalization.}
Consider a mesh $M_\theta$ with a neural texture parameterized by an MLP $\theta$ (This MLP could be either $\mathcal{F}_{\theta}$, $\mathcal{F}_{\phi}$, and $\mathcal{F}_{\psi}$ in \cref{subsec:visual-guidance}). We first render $M_\theta$ at $N$ different resolutions using a differentiable renderer $g$ to get multiple images $g(M_\theta) = \mathbf{x} = \{x^{1} \ldots x^{N}\}$ such that $x^{i}$ is the same resolution as stage $\phi^i$. For the base stage $\phi^1$, we perform standard SDS using \cref{eq:sds} on $x^1$ and prompt $y$ to get a gradient $\nabla_{x^1}$. For all stages $\phi^{i}$ for $i>1$, we sample two timesteps $t,s \sim \mathcal{U}(\{1,\ldots,T\})$, noise $\epsilon^i \sim \mathcal{N}(\mathbf{0},\mathbf{I})$ at the resolution of stage $\phi^i$, and noise $\epsilon^{i-1} \sim \mathcal{N}(\mathbf{0},\mathbf{I})$ at the resolution of stage $\phi^{i-1}$. Using timestep-dependent schedule coefficients $\alpha$ and $\sigma$, we compute a noisy image $z^{i}_t = \alpha_t x^i + \sigma_t \epsilon^{i}$ by applying a timestep-dependent scaling of $\epsilon^i$ to the image $x^i$. Similarly, we compute $z^{i-1}_s = \alpha_s x^{i-1} + \sigma_s \epsilon^{i-1}$ by applying a timestep-dependent scaling of $\epsilon^{i-1}$ to the image $x^{i-1}$. We then use $\phi^i$ to predict noise $\epsilon_{\phi^i}(z^{i}_t,t,z^{i-1}_s,s,y)$ conditioned on the noisy images, timesteps, and text prompt. Our gradient $\nabla_{x^i}$ for stage $\phi^{i}$ for $i>1$ is the difference between the predicted noise and the (higher-resolution) sampled noise $\epsilon^i$, weighted by the timestep-dependent function $w(t)$:
\begin{align}
    \nabla_{x^i} & \mathcal{L}_{CSD^i}(\phi^i, x^i, x^{i-1}, y) = \nonumber \\
    & w(t) (\epsilon_{\phi^i}(z^{i}_t,t,z^{i-1}_s,s,y) - \epsilon^i).
    \label{eq:csd-image-gradient}
\end{align}
With all gradients $\nabla_{x^1},\ldots,\nabla_{x^N}$ computed, we weight each gradient $\nabla_{x^i}$ with a user defined $\lambda^i$ to provide control over the impact of the supervision from each stage of the cascaded model. Thus our full gradient with respect to any given neural texture $\theta$ can be described by:
\begin{align}
    \nabla_{\theta} \mathcal{L}_{CSD} & (\boldsymbol{\phi}, \mathbf{x}=g(\theta), y) = \nonumber \\        
    & \lambda^1 \nabla_{x^1} \mathcal{L}_{SDS}(\phi^1, x^1, y) \frac{\partial x^1}{\partial \theta} \nonumber \\ 
    & + \sum_{i=2}^N \lambda^i \nabla_{x^i} \mathcal{L}_{CSD^i}(\phi^i, x^i, x^{i-1}, y) \frac{\partial x^i}{\partial \theta}.
    \label{eq:csd}
\end{align}
Note that just as in SDS~\cite{poole2022dreamfusion}, we can avoid computing the U-Net Jacobian term $\frac{\partial \epsilon_{\phi}(z^{i}_t, t, z^{i-1}_s, s, y)}{z^{i}_t}$ (not shown in \cref{eq:csd}) since each stage is entirely independent and our gradient is only with respect to the high-resolution image $x^i$. Thus, we directly apply $\lambda^i \nabla_{x^i}$ to the image $x^i$ without having to compute the costly backpropagation through the U-Net. Using the gradient $\nabla_\theta \mathcal{L}_{CSD}(\boldsymbol{\phi}, \mathbf{x}=g(\theta), y)$, we update the weights of our MLP $\theta$.

%% file: figures/overhead/gallery.tex
\begin{figure*}[t]
    \centering
    \newcommand{\one}{16}
    \newcommand{\two}{-2}
    \begin{overpic}[width=\linewidth]{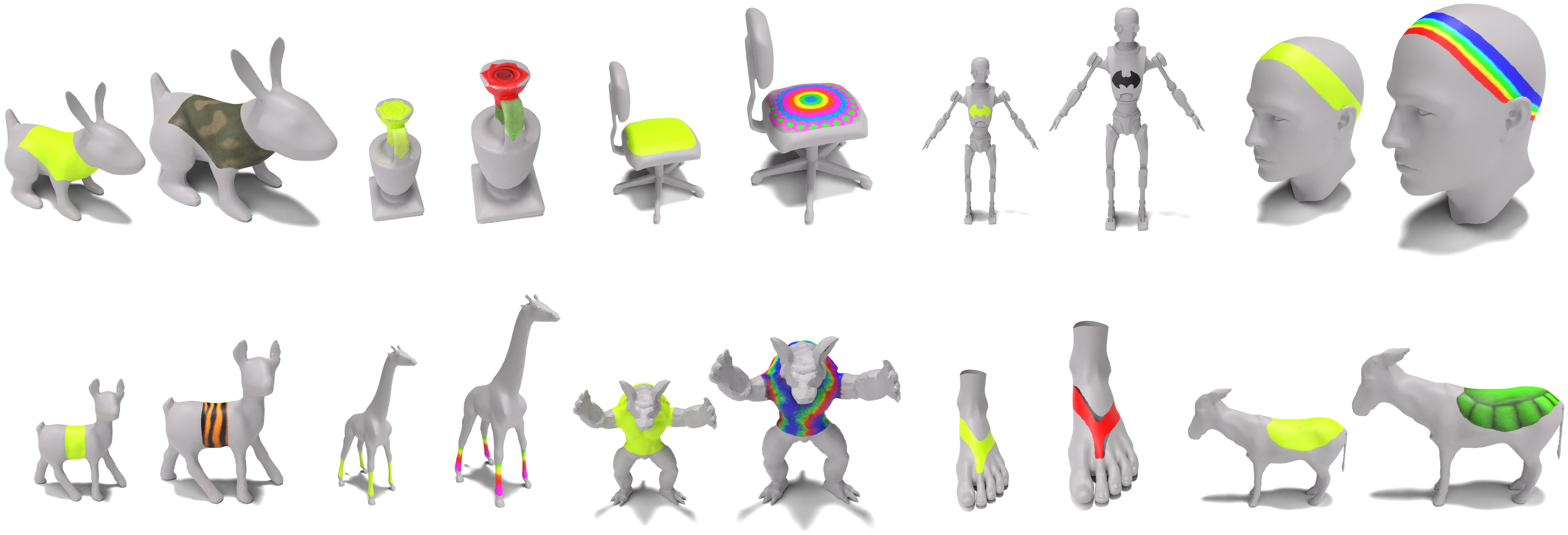}
    \put(5, \one){\textcolor{black}{Camo poncho}}
    \put(23, \one){\textcolor{black}{Beautiful rose}}
    \put(37, \one){\textcolor{black}{Colorful doily seat cushion}}
    \put(62, \one){\textcolor{black}{Batman emblem}}
    \put(81, \one){\textcolor{black}{Rainbow headband}}
    \put(4, \two){\textcolor{black}{Tiger stripe belt}}
    \put(20, \two){\textcolor{black}{Rainbow shinguards}}
    \put(42, \two){\textcolor{black}{Tie-dye shirt}}
    \put(61, \two){\textcolor{black}{Red flip flops}}
    \put(79, \two){\textcolor{black}{Green ridged turtle shell}}
    \end{overpic}
    \vspace{-2mm}
    \caption{\ourmethod{} produces highly detailed textures and localizations for a diverse range of meshes and prompts. Our method synthesizes meaningful local edits on shapes, demonstrating both global and local part-level understanding.}
    \label{fig:gallery}
\end{figure*}

%% file: figures/overhead/simultaneous.tex
\begin{figure}
    \centering
    \newcommand{\one}{-4}
    \begin{overpic}[width=\linewidth]{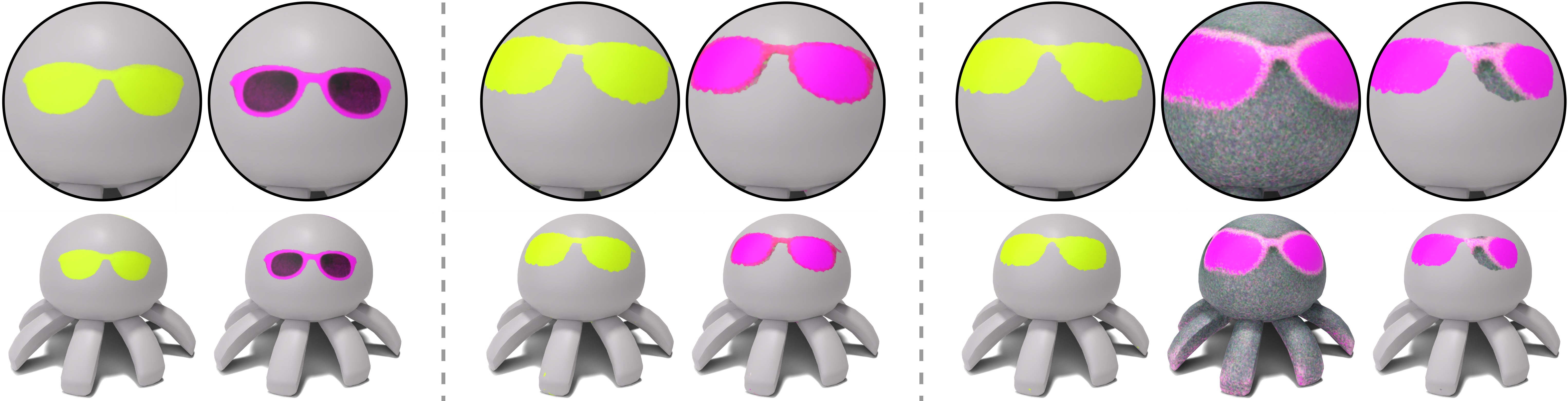}
    \put(3, \one){\textcolor{black}{Simultaneous}}
    \put(36, \one){\textcolor{black}{In series}}
    \put(70, \one){\textcolor{black}{Independent}}
    \end{overpic}
    \vspace{0.1mm}
    \caption{\textbf{Impact of simultaneous optimization}. Simultaneously optimizing the localization and texture (left) results in higher-detailed textures which effectively conform to the predicted localization. If we first optimize the localization, then optimize the texture within the localization region (\textit{in series}, middle), both the localization and texture are less detailed. Independent (right): if we optimize the localization independently (independent: left) and the texture independently (independent: middle), the texture does not align with the localization and thus the masked texture contains fringe artifacts (independent: right).}
    \label{fig:simultaneous}
\end{figure}

%% file: figures/overhead/csd_overview.tex
\begin{figure}
    \centering
    \includegraphics[width=\linewidth]{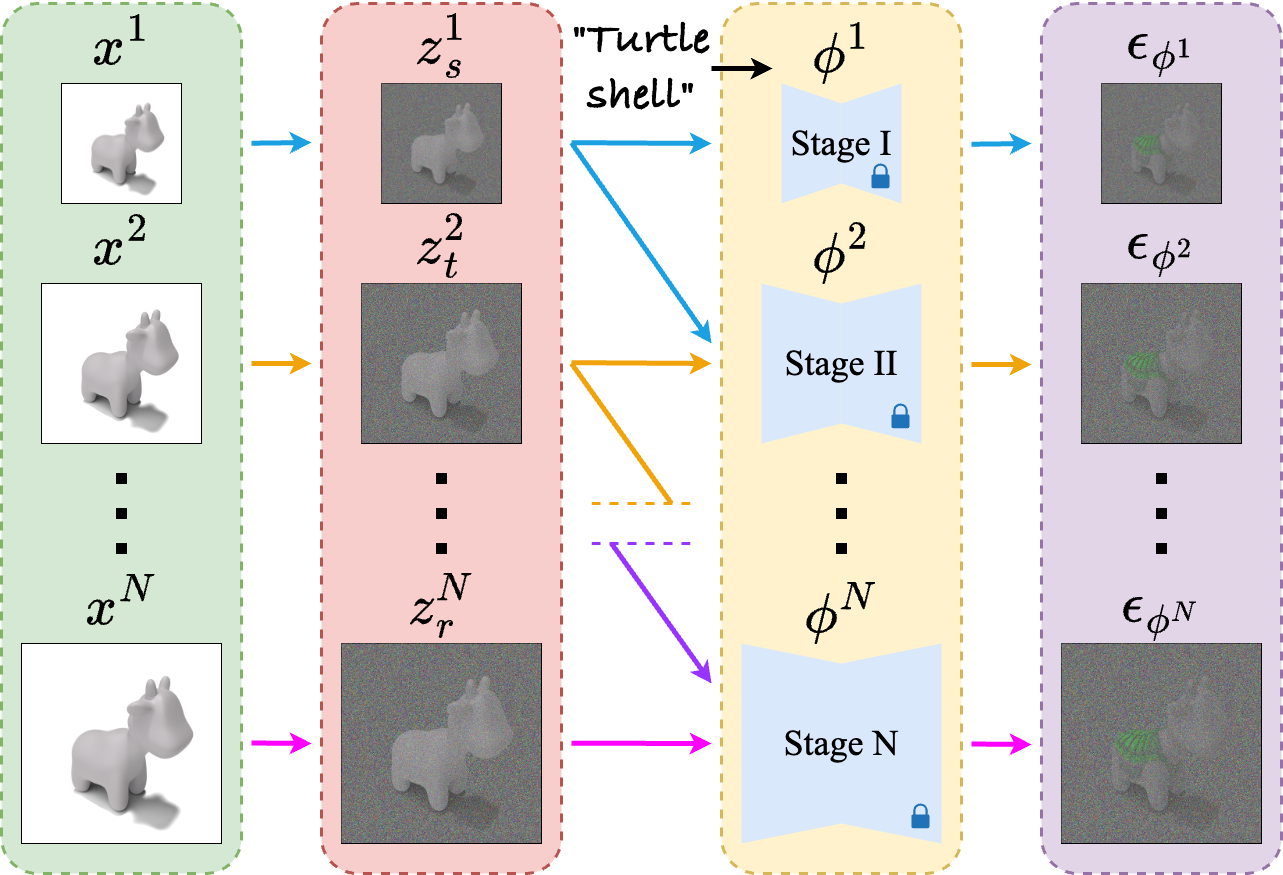}
    \caption{\textbf{Cascaded Score Distillation (CSD)}. We simultaneously distill scores across multiple stages of a cascaded diffusion model in order to leverage both the global awareness of the first stage and the higher level of detail contained in later stages. The difference between the predicted noise and sampled noise is the image gradient for each stage.}
    \label{fig:csd-overview}
\end{figure}

%% file: figures/overhead/csd_ablation.tex
\begin{figure}
    \centering
    \newcommand{\pl}{-5}
    \begin{overpic}[width=\linewidth]{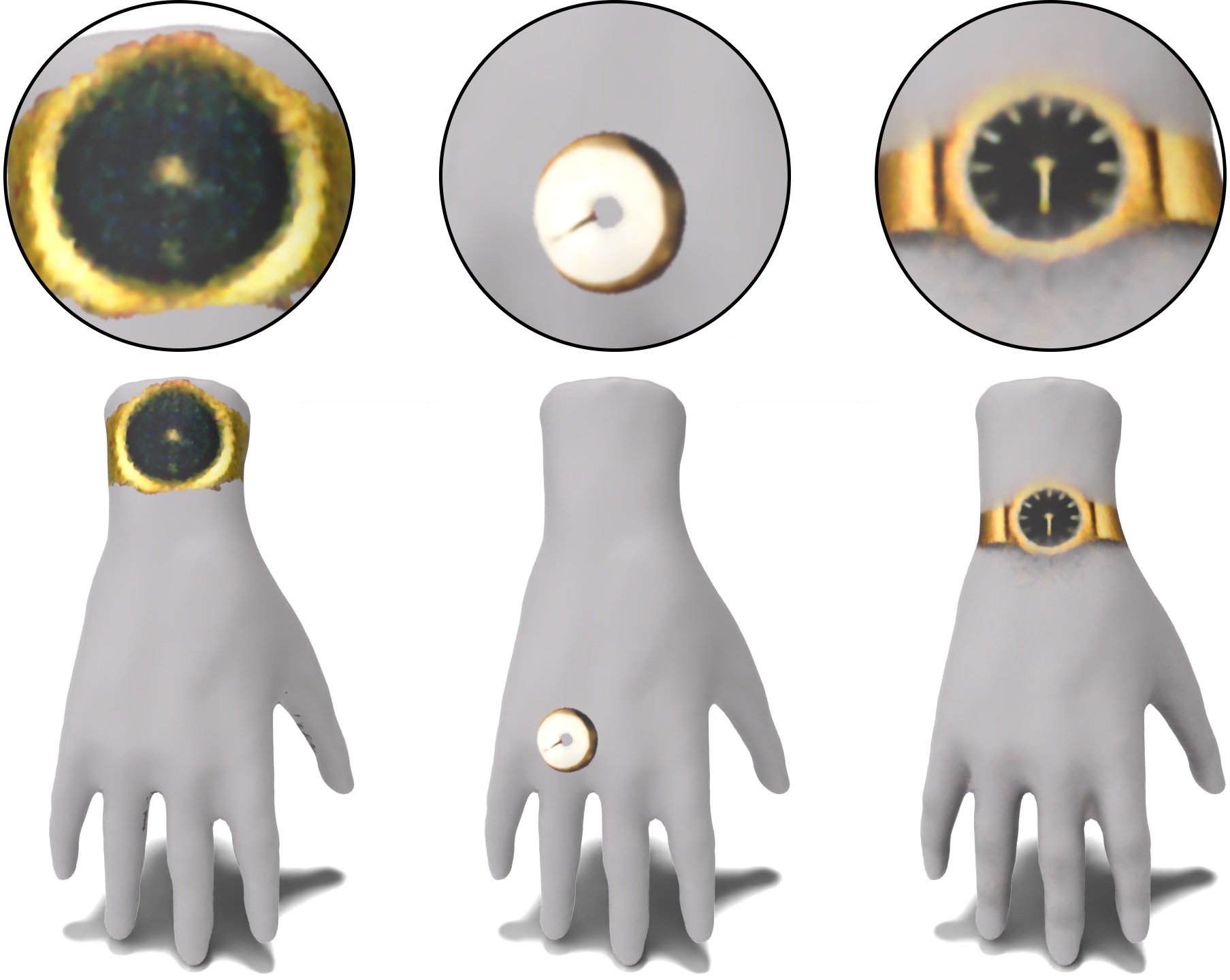}
    \put(4,  \pl){\textcolor{black}{Only stage 1}}
    \put(40,  \pl){\textcolor{black}{Only stage 2}}
    \put(82,  \pl){\textcolor{black}{CSD}}
    \end{overpic}
    \vspace{-2mm}
    \caption{\textbf{Impact of cascaded stages}. Different stages of the cascaded model provide different levels of granularity and global understanding. Using only the (low resolution) stage 1 model gives a low-resolution result in roughly the correct location. While the (high resolution) stage 2 model gives a high-resolution result, it is placed in the incorrect location. Our CSD simultaneously uses stage 1 and 2, resulting in a high-detailed texture in the appropriate location.}
    \label{fig:csd-ablation}
\end{figure}

%% file: figures/overhead/csd_control.tex
\begin{figure*}
    \centering
    \includegraphics[width=\linewidth]{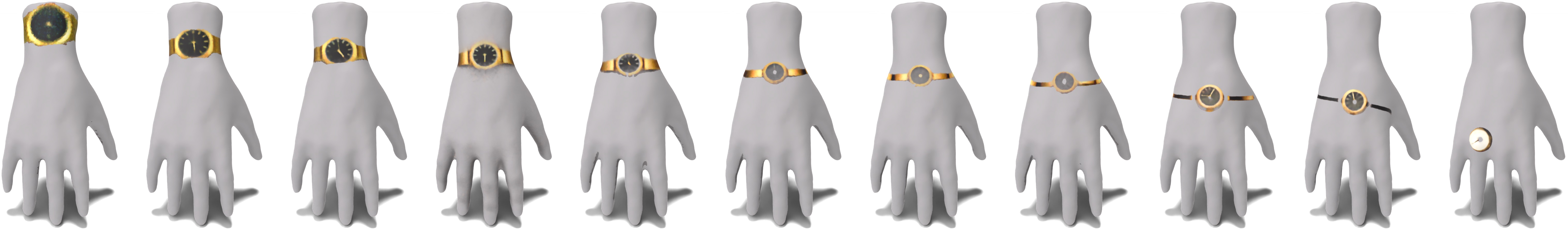}
    \caption{\textbf{Granular control with CSD}. Varying the weight between stage 1 and stage 2 results in control over the details and corresponding localization. Only using stage 1 (leftmost) is rather coarse; only using stage 2 (rightmost) is highly detailed with an incorrect localization. Increasing the stage 2 weight (moving left to right) progressively increases the detail and granularity of the supervision, enabling smooth and meaningful interpolation between stage 1 and 2. 
    }
    \label{fig:csd-control}
\end{figure*}

%% file: sec/4_experiments.tex
\input{figures/overhead/csd_sds_comp}
\input{figures/overhead/grid}
\section{Experiments}
In this section we demonstrate the capabilities of \ourmethod{} on a wide variety of meshes and prompts.

\smallskip
\noindent
\textbf{\ourmethod{} generality.} \ourmethod{} is capable of producing highly detailed localizations and textures on a diverse collection of meshes and prompts (\cref{fig:gallery}). Our method is not restricted to any category of meshes and we show results on organic meshes such as humanoids and animals as well as manufactured shapes such as household objects and furniture. Furthermore, our local textures can be specified with open vocabulary text descriptions and are not limited to any predefined categories or constraints. This includes ``out-of-domain" local textures such as shinguards on a giraffe or a shirt on a goat which are not naturally seen in the context of these objects, yet are precisely placed in semantically meaningful locations with highly detailed textures.

\smallskip
\noindent
\textbf{\ourmethod{} precision and composition.} \ourmethod{} produces precise localizations and highly-detailed textures that effectively adhere to these predicted localizations (see \cref{fig:composite}). The tight coupling between the localization and texture (see the gold chain necklace in \cref{fig:teaser}) enables seamless composition of multiple local textures simultaneously on the same mesh without any layering artifacts. For example, the sharp localization boundary of the ``Tie-dye apron'' (in \cref{fig:composite}) allows us to composite this local texture on top of other textures such as the superman emblem without affecting the emblem's appearance in regions outside the boundary of apron texture.

\smallskip
\noindent
\textbf{\ourmethod{} specificity and robustness.} 
\ourmethod{} produces accurate and high resolution local edits that closely adhere to the text-specification (see \cref{fig:grid}). Our method's fine-grained results contain intricate details (\ie the badge on ``Barcelona jersey'') and reflect the subtle differences in the text prompts (\ie the ``cape'' on the dog is more tapered than the boxier ``poncho''). This specificity allows us to produce many diverse and distinct local styles. We show multiple local edits on the same mesh for multiple different meshes, demonstrating the robustness of our method with respect to both the prompts and meshes.

\smallskip
\noindent
\textbf{Impact and granular control of CSD.}
Our cascaded score distillation (CSD) simultaneously distills scores at multiple resolutions in a cascaded fashion. We observe that different stages of the cascaded diffusion model give different levels of granularity and global understanding (\cref{fig:csd-ablation}). Using only the (low resolution) stage 1 model is equivalent to SDS and while this produces an accurate localization and coherent texture, the result is low-resolution (see \cref{fig:csd-sds-comp}). Conversely, using only the (high resolution) stage 2 model gives a high-resolution result, but often fails to properly localize the edit leading to unrealistic results. Our CSD simultaneously combines the supervision from stages 1 and 2, resulting in a highly-detailed texture in the appropriate location. Increasing the stage 2 weight (moving left to right in \cref{fig:csd-control}) progressively increases the detail and granularity of the supervision, demonstrating smooth and intuitive interpolation between stage 1 and 2. Thus, a user may choose to vary the weight between stage 1 and stage 2 to intuitively control the influence of each stage and in turn the granularity and global understanding of the supervision (\cref{fig:csd-control}).

\input{figures/overhead/loss_ablation}

\smallskip
\noindent
\textbf{Simultaneous localization and texture.}
We demonstrate the importance of simultaneously optimizing the localization and texture region in tandem in \cref{fig:simultaneous}. We observe that simultaneous optimization results in highly detailed textures which effectively conform to the predicted localization regions (\cref{fig:simultaneous}, left). The predicted localization region is sharp and intricate. Alternatively, we optimize the localization region first (\cref{fig:simultaneous}, middle). Then, we use the predicted localization to learn a texture which is confined to the (pre-computed) localization region. We observe that both the texture and localizations are of lower-quality. The texture is less detailed, and the localization region is less intricate. Finally, it is possible to learn the texture and localization region completely independently from one another (\cref{fig:simultaneous}, independent). This results in a texture (\cref{fig:simultaneous} independent, middle) that is completely decoupled from the localization (\cref{fig:simultaneous} independent, left) region. When masking the texture with the localization region (\cref{fig:simultaneous} independent, right) we observe a misaligned texture with fringe artifacts.

\smallskip
\noindent
\textbf{Background loss ablation.}
We opt to learn the background texture in a separate map from the desired texture map (middle and bottom branches of \cref{fig:overview}) enabling our method to produce accurate localizations with tightly coupled and detailed textures (\cref{fig:background-ablation}, left). If we instead remove the explicit background texture map (and MLP) and allow the background to be predicted using the same texture map as the main edit texture $T_{map}$, (by applying background loss to $T_{map}$ masked with the inverse of the localization mask), then the localization and texture become misaligned
(\cref{fig:background-ablation}, middle). When the edit and background share the same texture map, if the localization region expands during training to include features that had been considered background, the edit texture may retain these features of the object (rather than expand the edit texture features to fill the localization) and the localization may continue to expand. If we instead remove the background loss entirely, this also produces undesirable results (\cref{fig:background-ablation}, right). Specifically, we observe extra elements incorporated in additional localization regions that contain characteristics of the input shape (\eg bill on a duck).

%% file: figures/overhead/csd_sds_comp.tex
\begin{figure}[b]
    \centering
    \newcommand{\pl}{-4}
    \begin{overpic}[width=\linewidth]{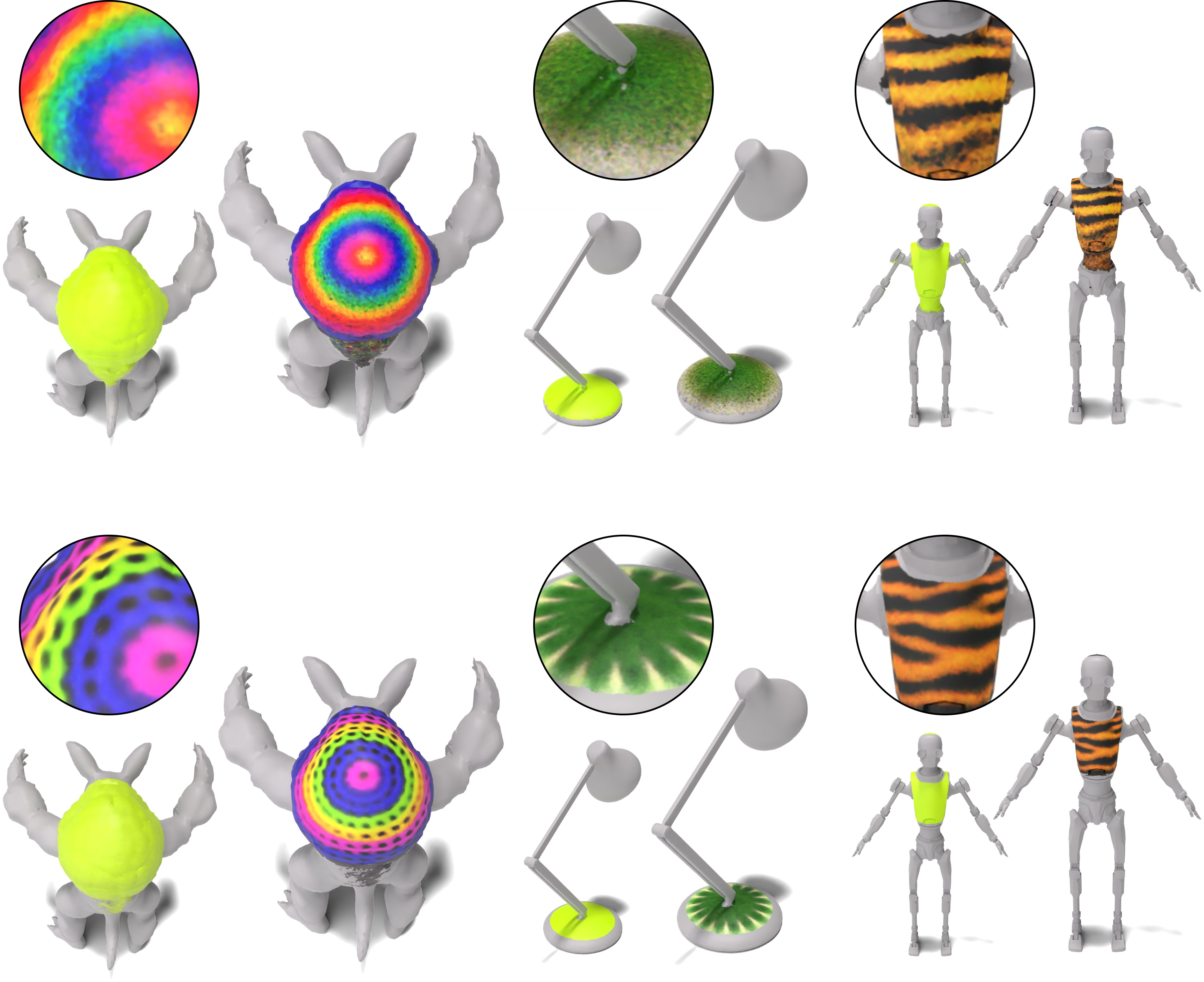}
    \put(47,  41){\textcolor{black}{SDS}}
    \put(47,  \pl){\textcolor{black}{CSD}}
    \end{overpic}
    \vspace{-2mm}
    \caption{\textbf{Importance of super-resolution stage in CSD.} Using stage 1 only (equivalent to SDS) lacks fine-grained details. Incorporating the second super-resolution cascaded stage from our CSD increases the resolution and detail. Input text prompts (from left to right): Colorful crochet shell, Cactus base, Tiger stripe shirt.}
    \label{fig:csd-sds-comp}
\end{figure}

%% file: figures/overhead/grid.tex
\begin{figure*}
    \centering
    \newcommand{\one}{20}
    \newcommand{\two}{-2}
    \begin{overpic}[width=\linewidth]{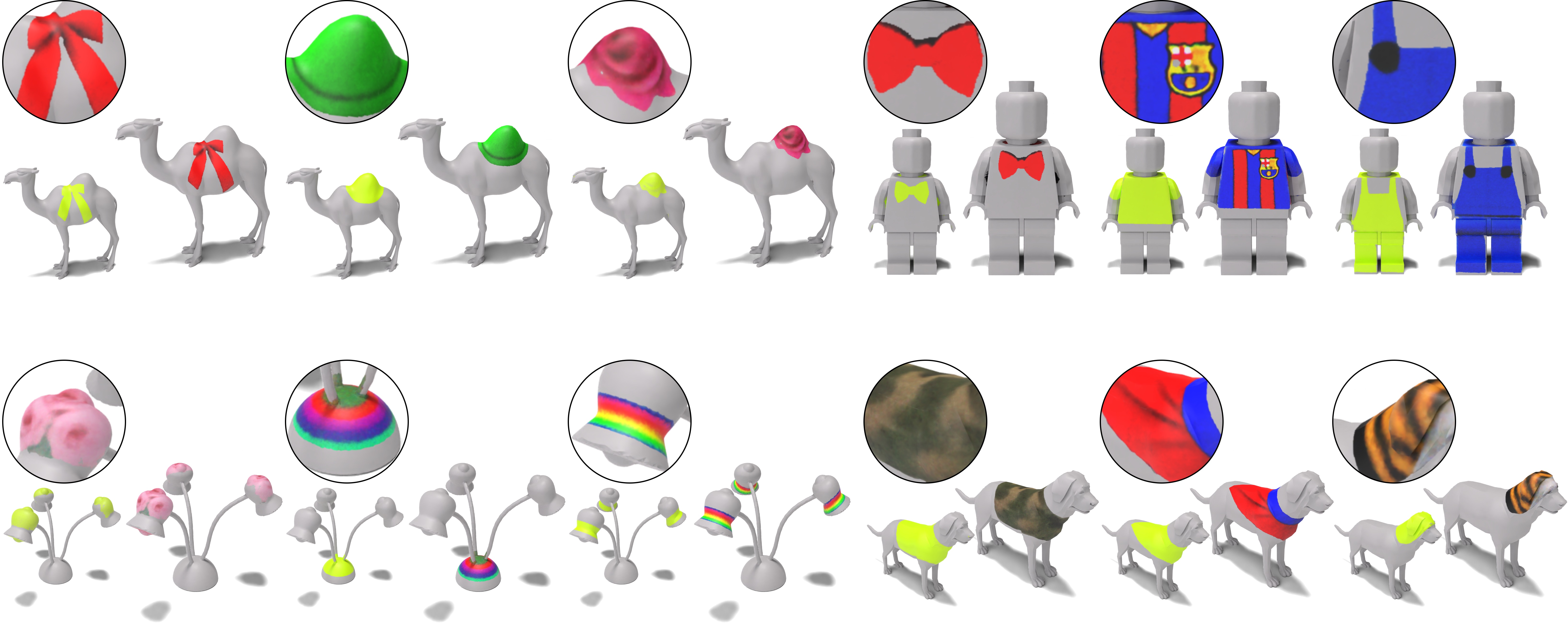}
    \put(5,  \one){\textcolor{black}{Red bow tie}}
    \put(23,  \one){\textcolor{black}{Turtle shell}}
    \put(37,  \one){\textcolor{black}{Beautiful rose hump}}
    \put(57,  \one){\textcolor{black}{Red bow tie}}
    \put(70,  \one){\textcolor{black}{Barcelona jersey}}
    \put(86,  \one){\textcolor{black}{Denim overalls}}
    \put(2,  \two){\textcolor{black}{Beautiful roses}}
    \put(17,  \two){\textcolor{black}{Colorful crochet base}}
    \put(38,  \two){\textcolor{black}{Rainbow headband}}
    \put(56,  \two){\textcolor{black}{Camo poncho}}
    \put(71,  \two){\textcolor{black}{Superhero cape}}
    \put(86,  \two){\textcolor{black}{Tiger stripe hat}}
    \end{overpic}
    \vspace{-2mm}
    \caption{\ourmethod{} is capable of producing a variety of local textures on the same mesh. Each result contains an accurate localization map (to specify the edit region) and a texture map that conforms to it.}
    \label{fig:grid}
\end{figure*}
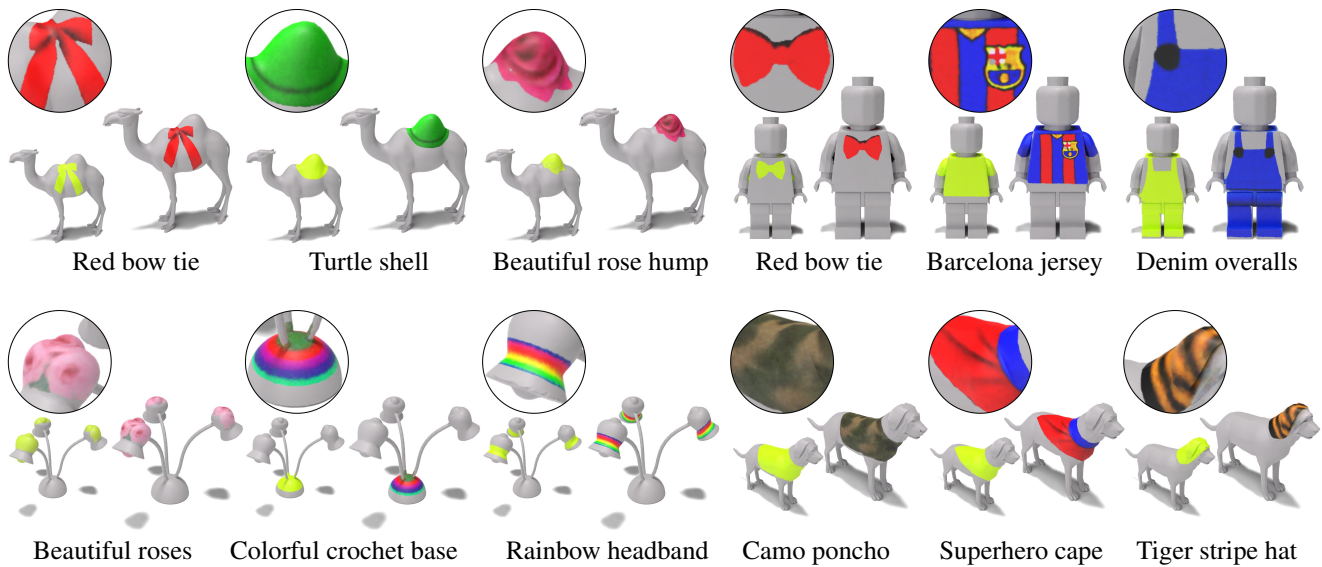
 

%% file: figures/overhead/loss_ablation.tex
\begin{figure}[b]
    \centering
    \newcommand{\pl}{-5}
    \begin{overpic}[width=\linewidth]{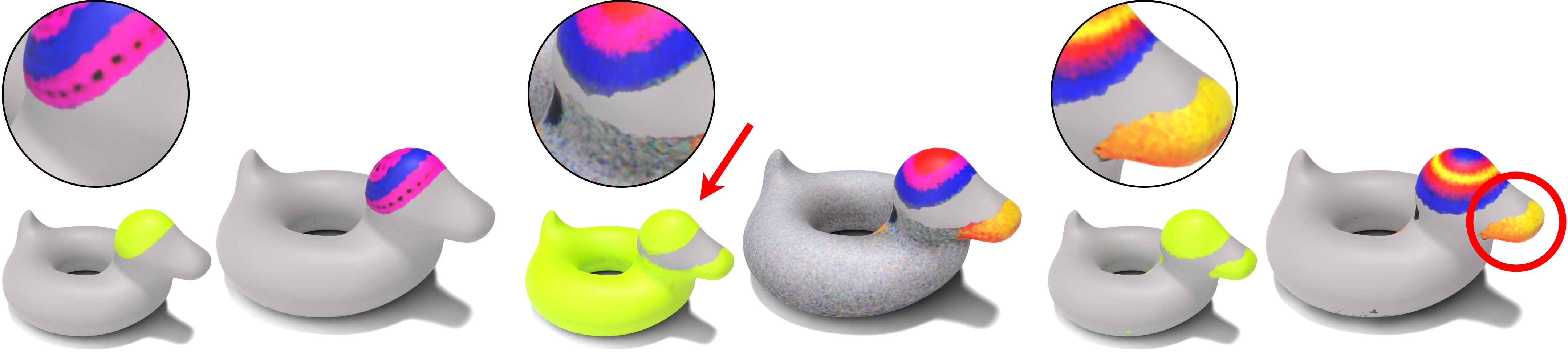}
    \put(13,  \pl){\textcolor{black}{Ours}}
    \put(32,  \pl){\textcolor{black}{w/o separate MLP}}
    \put(67,  \pl){\textcolor{black}{w/o background loss}}
    \end{overpic}
    \vspace{-2mm}
    \caption{\textbf{Background loss ablation}. Our method (left) produces an accurate localization and texture that are highly detailed and tightly coupled. Removing the background MLP $\mathcal{F}_{\psi}$ and instead learning the background through the main texture map $T_{map}$ leads to poor localization which also degrades the texture (middle). Removing the background loss completely leads to the incorporation of superfluous elements from the input model (\ie a bill on a duck) into the localization (right).}
    \label{fig:background-ablation}
\end{figure}

%% file: sec/5_conclusion.tex
\section{Conclusion}
We presented \ourmethod, a technique that produces highly detailed texture maps on meshes which effectively adhere to a predicted localization region. Our system is capable of \emph{hallucinating} a non-obvious local texture on a wide variety of meshes (such as heart-shaped sunglasses on a cow). Our localizations are detailed and accurate, enabling seamless post-processing (such as compositing textures without unwanted fringe).

We proposed cascaded score distillation, a technique capable of extracting supervision signals from multiple stages of a cascaded diffusion model. We observe that each stage controls different amounts of detail and global understanding. Further, varying the weights for each stage provides control over the resulting local textures. We show the effectiveness of CSD to locally texture meshes; however, CSD is general and can be applied to other domains (such as images, videos, and alternative 3D representations).

In the future, we are interested in extending the localized editing capabilities beyond texturing (such as deformations, normal maps, materials, and more). Another avenue for future work is learning to \textit{co-texture} multiple shapes using the same local texture map, which may provide correspondences between shapes.

\section{Acknowledgments}

We thank the University of Chicago for providing the AI cluster resources, services, and the professional support of the technical staff. This work was also supported in part by gifts from Snap Research, Adobe Research, Google Research, BSF grant 2022363, and NSF grants 2304481 and 2241303. Finally, we would like to thank Brian Kim, Jack Zhang, Haochen Wang, and the members of 3DL and PALS for their thorough and insightful feedback on our work.